\documentclass[twocolumn,prl,a4paper,superscriptaddress,showpacs,preprintnumbers,tightenlines]{revtex4}

\usepackage{amsmath}
\usepackage{graphicx}
\usepackage{epsfig}
\usepackage{axodraw}

\begin{document}

\preprint{\tighten\vbox{\hbox{\hfil Belle Preprint 2004-15}}}
\preprint{\tighten\vbox{\hbox{\hfil KEK   Preprint 2004-19}}}

\title{Measurement of Branching Fraction and $CP$ Asymmetry in
\mbox{$\boldmath{B^+\rightarrow \rho^+ \pi^0}$}}

\affiliation{Budker Institute of Nuclear Physics, Novosibirsk}
\affiliation{Chiba University, Chiba}
\affiliation{Chonnam National University, Kwangju}
\affiliation{University of Cincinnati, Cincinnati, Ohio 45221}
\affiliation{Gyeongsang National University, Chinju}
\affiliation{University of Hawaii, Honolulu, Hawaii 96822}
\affiliation{High Energy Accelerator Research Organization (KEK), Tsukuba}
\affiliation{Hiroshima Institute of Technology, Hiroshima}
\affiliation{Institute of High Energy Physics, Chinese Academy of Sciences, Beijing}
\affiliation{Institute of High Energy Physics, Vienna}
\affiliation{Institute for Theoretical and Experimental Physics, Moscow}
\affiliation{J. Stefan Institute, Ljubljana}
\affiliation{Kanagawa University, Yokohama}
\affiliation{Korea University, Seoul}
\affiliation{Kyungpook National University, Taegu}
\affiliation{Swiss Federal Institute of Technology of Lausanne, EPFL, Lausanne}
\affiliation{University of Ljubljana, Ljubljana}
\affiliation{University of Maribor, Maribor}
\affiliation{University of Melbourne, Victoria}
\affiliation{Nagoya University, Nagoya}
\affiliation{Nara Women's University, Nara}
\affiliation{National United University, Miao Li}
\affiliation{Department of Physics, National Taiwan University, Taipei}
\affiliation{H. Niewodniczanski Institute of Nuclear Physics, Krakow}
\affiliation{Nihon Dental College, Niigata}
\affiliation{Niigata University, Niigata}
\affiliation{Osaka City University, Osaka}
\affiliation{Osaka University, Osaka}
\affiliation{Panjab University, Chandigarh}
\affiliation{Peking University, Beijing}
\affiliation{Princeton University, Princeton, New Jersey 08545}
\affiliation{Saga University, Saga}
\affiliation{University of Science and Technology of China, Hefei}
\affiliation{Seoul National University, Seoul}
\affiliation{Sungkyunkwan University, Suwon}
\affiliation{University of Sydney, Sydney NSW}
\affiliation{Tata Institute of Fundamental Research, Bombay}
\affiliation{Toho University, Funabashi}
\affiliation{Tohoku Gakuin University, Tagajo}
\affiliation{Tohoku University, Sendai}
\affiliation{Department of Physics, University of Tokyo, Tokyo}
\affiliation{Tokyo Institute of Technology, Tokyo}
\affiliation{Tokyo Metropolitan University, Tokyo}
\affiliation{Tokyo University of Agriculture and Technology, Tokyo}
\affiliation{University of Tsukuba, Tsukuba}
\affiliation{Virginia Polytechnic Institute and State University, Blacksburg, Virginia 24061}
\affiliation{Yonsei University, Seoul}

  \author{J.~Zhang}\affiliation{High Energy Accelerator Research Organization (KEK), Tsukuba} 
  \author{K.~Abe}\affiliation{High Energy Accelerator Research Organization (KEK), Tsukuba} 
  \author{K.~Abe}\affiliation{Tohoku Gakuin University, Tagajo} 
  \author{T.~Abe}\affiliation{High Energy Accelerator Research Organization (KEK), Tsukuba} 
  \author{H.~Aihara}\affiliation{Department of Physics, University of Tokyo, Tokyo} 
  \author{Y.~Asano}\affiliation{University of Tsukuba, Tsukuba} 
  \author{V.~Aulchenko}\affiliation{Budker Institute of Nuclear Physics, Novosibirsk} 
  \author{T.~Aushev}\affiliation{Institute for Theoretical and Experimental Physics, Moscow} 
  \author{T.~Aziz}\affiliation{Tata Institute of Fundamental Research, Bombay} 
  \author{S.~Bahinipati}\affiliation{University of Cincinnati, Cincinnati, Ohio 45221} 
% ...
\collaboration{The Belle Collaboration}

\begin{abstract}
We report a measurement of the branching fraction for the decay
$B^+ \to \rho^+\pi^0$ based on a 140~${\rm fb}^{-1}$ data sample
collected with the Belle detector at the KEKB asymmetric $e^+e^-$
collider.
We measure the branching fraction ${\cal B}(B^+ \to
\rho^+\pi^0)=(13.2\pm 2.3({\rm stat.})^{+1.4}_{-1.9}({\rm
  sys.}))\times 10^{-6}$, and the $CP$-violating asymmetry ${\cal
  A}_{CP}(B^\mp \to \rho^\mp\pi^0)=0.06\pm0.19({\rm
  stat.})^{+0.04}_{-0.06}({\rm sys.})$.

\end{abstract}


\maketitle
Recent precise measurements of $\sin 2\phi_1$ \cite{Belle-sin2phi1,
  Babar-sin2phi1} confirm the prediction of the Kobayashi-Maskawa
model ~\cite{ckm} for $CP$ violation. It is of great importance to test
this theory further with complementary measurements, such as
those of the other unitarity triangle angles $\phi_2$ and
$\phi_3$~\cite{phi1-3}.

At the quark level, the decays $B \to \rho \pi$ occur via  $b\to u$
tree diagrams and can be used to measure $\phi_2$.
However, because of the presence of $b\to d$ penguin (loop) diagrams,
a model independent extraction of $\phi_2$ from time-dependent
$CP$-asymmetry measurements requires an isospin analysis of the decay
rates of all the $\rho \pi$ decay modes~\cite{iso_ana}.
The decay channels $B^+ \to \rho^0 \pi^+$ ~\cite{charge_conjugate} and
$B^0 \to \rho^{\pm} \pi^{\mp}$ have already been
measured~\cite{current_rhopi}.
Evidence for the $B^0 \to \rho^0 \pi^0$ mode, which is expected to be
small, has been reported by Belle~\cite{jasna}
with a rate higher than an upper bound from
Babar~\cite{BaBar_rhoppi0}.
The remaining decay mode, $B^+ \to \rho^+ \pi^0$, has two neutral pions
in the final state that make its measurement an experimental
challenge. Recently, the BaBar group reported the observation of this
mode~\cite{BaBar_rhoppi0}.

In this Letter, we report measurements of the branching fraction and
the $CP$-violating charge asymmetry for the $B^+\to \rho^+\pi^0$ decay
mode.
The results are based on a 140~fb$^{-1}$ data sample containing
$152.0\times 10^{6}$ $B$ meson pairs collected with
the Belle detector at the KEKB asymmetric-energy $e^+e^-$
collider~\cite{KEKB} operating at the $\Upsilon(4S)$ resonance
($\sqrt{s} = 10.58$~GeV). The production rates of $B^+B^-$ and
$B^0\overline{B}{}^0$ pairs are assumed to be equal.

The Belle detector is a large-solid-angle magnetic spectrometer that
consists of a three-layer silicon vertex detector, a 50-layer
central drift chamber (CDC), an array of aerogel threshold
\v{C}erenkov counters (ACC), a barrel-like arrangement of
time-of-flight scintillation counters (TOF), and an electromagnetic
calorimeter comprised of CsI(Tl) crystals (ECL)  located inside a
superconducting solenoid coil that provides a 1.5~T magnetic field.
An iron flux-return located outside of the coil is instrumented to
detect $K_L$ mesons and to identify muons.
The detector is described in detail elsewhere \cite{Belle}.

For charged pion and kaon identification, specific ionization
measurements ($dE/dx$) from the CDC are combined with the responses of
the ACC and TOF systems to form likelihoods ${L}_{\pi}$ and ${L}_K$. We
distinguish pions from kaons by applying selection requirements on the
likelihood ratio, ${L}_\pi/({L}_\pi+{L}_K)$. 
Similarly, electrons are identified by means of a likelihood based on
ECL measurements, $dE/dx$ information from the CDC, and the responses
of the ACC.

The final state for the signal consists of a charged pion track and
two $\pi^0\to\gamma\gamma$ candidates.
The charged track is required to have a transverse momentum $p_T>0.1$
GeV$/c$ and to be consistent with an origin within $0.1~{\rm cm}$ in
the radial direction and $5~\rm{cm}$ along the beam direction of the
interaction point (IP).
In addition, the charged track is required to be positively identified
as a pion, and not be consistent with the electron hypothesis.
Candidate $\pi^0$ mesons are reconstructed from pairs of photons that
have an invariant mass within $\pm 3\sigma$ of the nominal $\pi^0$
mass,
where the photons are assumed to originate from the IP, and the
$\pi^0$ resolution $\sigma$ varies in the range $5.3~{\rm MeV}$ -- $7.0
{~\rm MeV}$ depending on its momentum.
The energy of each photon in the laboratory frame is required to be
greater than 50 MeV for the ECL barrel region 
($32^\circ< \theta <129^\circ$) and 100 MeV for the ECL endcap
regions ($17^\circ<\theta<32^\circ$ or
$129^\circ<\theta<150^\circ$), where $\theta$ denotes the polar
angle of the photon with respect to the beam line. 
The $\pi^0$ candidates are kinematically constrained to the nominal
$\pi^0$ mass.
In order to reduce the combinatorial background, we only accept $\pi^0$
candidates with momenta $p_{\pi^0}>0.35$ GeV$/c$ in the $e^+e^-$
center-of-mass system (CMS).
We select $\rho^+ \to \pi^+\pi^0$ decay candidates with invariant
masses in the range $0.62{~\rm GeV}/c^2<M(\pi^+\pi^0)<0.92{~\rm GeV}/c^2$.

$B^+\to \rho^+\pi^0$ candidates are identified using the 
beam-constrained mass $M_{\rm bc}\equiv\sqrt{E_{\rm beam}^2-p_B^2}$, 
and the energy difference $\Delta E\equiv E_B-E_{\rm beam}$, 
where $E_{\rm beam}$ is the CMS beam energy, and $p_B$ and $E_B$ are
the CMS momentum and energy, respectively, of the $B$ candidate.
The $\Delta E$ distribution has a tail on  the lower side caused by
incomplete longitudinal containment of electromagnetic showers in the
CsI crystals.
We accept events in the region $M_{\rm bc}>5.2 {~\rm GeV}/c^2$,
$-0.4{~\rm GeV}<\Delta E<0.4{~\rm GeV}$ and define a signal region as
$5.27{~\rm GeV}/c^2<M_{\rm bc}<5.29{~\rm GeV}/c^2$ 
and $-0.20{~\rm GeV}<\Delta E<0.07{~\rm GeV}$.

The continuum process $e^+e^- \to q \bar{q}$ ($q=u, d, s, c$) is the
main source of background and must be strongly suppressed. 
One method of discriminating the signal from the background is based on
the event topology, which tends to be isotropic for $B\bar B$ events
and jet-like for $q\bar q$ events.  Another discriminating
characteristic is $\theta_B$, the CMS polar angle of the $B$ flight
direction.
$B$ mesons are produced with a $1-\cos^2\theta_B$ distribution
while continuum background events tend to be uniform in $\cos\theta_B$.
We require $|\cos\theta_{\rm thr}|<0.8$, where $\theta_{\rm thr}$
is the angle between the thrust axis of the candidate tracks plus
neutrals and that of the remaining tracks in the event.  This
distribution is flat for signal events and peaked at $\cos\theta_{\rm
  thr}=\pm 1$ for continuum backgrounds.
We use Monte Carlo (MC) simulated signal and continuum
events to form a
Fisher discriminant based on modified Fox-Wolfram moments~\cite{shlee}
that are verified to be uncorrelated with $M_{\rm bc}$, $\Delta E$ and
variables considered later in the analysis.
Probability density functions (PDFs) derived from the
Fisher discriminant and the $\cos\theta_B$ distributions are
multiplied to form likelihood functions for signal (${\cal L}_{s}$)
and continuum (${\cal L}_{q\bar q}$);
these are combined into a likelihood ratio 
${\cal R}_s={\cal L}_{s}/({\cal L}_{s}+{\cal L}_{q\bar q})$.
Additional discrimination is provided by the $b$-flavor
tagging parameter $r$, which ranges from 0 to 1 and is
a measure of the likelihood that the $b$ flavor 
of the accompanying $B$ meson is correctly assigned
by the Belle flavor-tagging algorithm~\cite{Belle-sin2phi1}. 
Events with high values of $r$ are well-tagged and are less
likely to originate from continuum production.
We define a multi-dimensional likelihood ratio $MDLR= {\cal 
L}_{s}^{MDLR}/({\cal L}_{s}^{MDLR}+ {\cal 
L}_{q\bar q}^{MDLR})$, where ${\cal L}_{s}^{MDLR}$
denotes the likelihood determined by the $r$-${\cal R}_s$
distribution for signal and ${\cal L}_{q\bar q}^{MDLR}$
is that for the continuum background. We achieve continuum suppression
by requiring ${MDLR}>0.9$.  In this way we reject 99\% of the
continuum background while retaining 30\% of the $B^+ \to \rho^+\pi^0$
signal.

Since $B\to \rho^+\pi^0$ is a ${\rm pseudoscalar} 
\to {\rm vector}+{\rm pseudoscalar}$ process, the $\rho$ helicity
angle $\theta_{\rm hel}$, 
defined as the angle between an axis anti-parallel to the $B$ flight
direction and the $\pi^+$ flight direction in the $\rho$ rest frame, 
has a $\cos^2\theta_{\rm hel}$ distribution.
We require  $|\cos \theta_{\rm hel}|>0.3$ for further background
suppression. 

In the $M_{\rm bc}$-$\Delta E$ signal region, about 8\% of the events
have multiple candidates. We choose the candidate that has the
minimum sum of $\chi^2$ for the mass constrained $\pi^0$ fits;  in
cases where candidates have the same $\chi^2$, we select the candidate
with the largest Fisher discriminant.
The MC-determined efficiency with all selection criteria imposed is
found to be 4.4\%.

Backgrounds from $B$ decays are investigated with MC simulation.
For $b\to c$ decay processes, no signal-like peak is found in
either the $M_{\rm bc}$ or $\Delta E$ distribution. 
Among the much rarer charmless decays, the dominant backgrounds are
from $B^0 \to \rho^+\rho^-$ decays, which populate the negative
$\Delta E$ region, and  $B^0 \to \pi^0\pi^0$ decays, which populate
the positive $\Delta E$ region.  
Monte Carlo studies indicate that potential backgrounds from $B^+\to
a_1^+\pi^0$, $a_1^+\to \rho^+\pi^0$ have $\Delta E$ and $M_{\rm bc}$
distributions similar to those for $B^0 \to \rho^+\rho^-$
decays and are accounted for by the latter component of the fit.
In addition, the contamination from other possible rare $B$ decays
is taken into account in the systematic error.

We extract the $B^+ \to \rho^+\pi^0$ signal yield by applying an
extended unbinned maximum-likelihood fit to the two-dimensional 
$M_{\rm bc}$-$\Delta E$ distribution.
The fit includes components for signal plus backgrounds from continuum
events, $b\to c$ decays, $B^0 \to \rho^+\rho^-$ and $B^0\to
\pi^0\pi^0$.
The PDFs for signal, $B^0\to \rho^+\rho^-$ and $B^0\to \pi^0\pi^0$ are
modeled by smoothed two-dimensional histograms obtained from large MC
samples. 
The signal PDF is adjusted to account for small differences observed
between data and MC for high-statistics modes containing $\pi^0$'s,
i.e.,
$B^+ \to {\overline D}{}^0(K^+\pi^-\pi^0) \pi^+$ for $M_{\rm
  bc}$, and $D^0 \to\pi^0\pi^0$ for $\Delta E$, where we use $\pi^0$
mesons in similar momentum ranges to those from $B^+\to \rho^+\pi^0$
decay.
The continuum PDF is described by a product of a threshold (ARGUS)
function~\cite{ARGUS} for $M_{\rm bc}$ and a first-order
polynomial for $\Delta E$, with shape parameters allowed to be free.
Background from generic $b\to c$ decays is represented by an ARGUS
function for $M_{\rm bc}$ and a third-order polynomial for
$\Delta E$ with shape parameters determined from MC.
In the fit, all normalizations are allowed to float, except for the
$\pi^0\pi^0$ component, which is fixed at a MC-determined value based
on recent Belle~\cite{shlee} and BaBar~\cite{BaBar_pi0pi0} measurements.

\begin{figure}[htb]
\begin{center}
\includegraphics[width=5.6cm,clip]{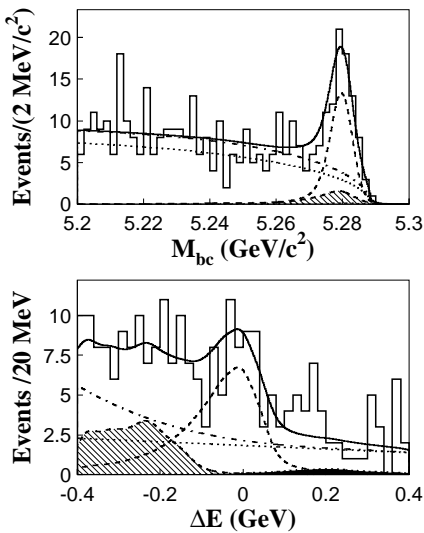}
\caption{The upper plot is the $M_{\rm bc}$ projection for events in
  the $\Delta E$ signal region $-0.2{~\rm GeV}<\Delta E<0.07{~\rm
  GeV}$; 
  the lower plot is the $\Delta E$  projection for events in 
  the $M_{\rm bc}$ signal region  $5.27{~\rm GeV}/c^2<M_{\rm
  bc}<5.29{~\rm GeV}/c^2$. 
  The solid curve shows the results of the fit.
  The signal component is shown as a dashed line.
  The continuum background is shown as a dotted line.
  The sum of $b\to c$ and continuum components is shown as a
  dot-dashed line.
  The hatched (dark) histogram represents the $B^0\to \rho^+\rho^-$
  ($B^0\to \pi^0\pi^0$) background.}
\label{fitdata_proj}
\end{center}
\end{figure}

Figure~\ref{fitdata_proj} shows the final event sample and fit results.
The six-parameter (four normalizations plus two shape parameters for
continuum) fit gives a signal yield of $87 \pm 15$ events. 
The statistical significance of the signal, 
defined as $\sqrt{-2\ln({\cal L}_0/{\cal L}_{\rm max})}$,
where ${\cal L}_{\rm max}$ is the likelihood value at the best-fit
signal  yield and ${\cal L}_{0}$ is the value with the signal yield
set to zero, is 8.1$\sigma$.
The level of the $B^0\to \rho^+\rho^-$ background determined from the
fit is in good agreement with MC-expectations based on a recent
measurement of the branching fraction for this decay
mode~\cite{BaBar_rhoprhom}.

\begin{figure}[htbp]
\begin{center}
\includegraphics[width=5cm,clip]{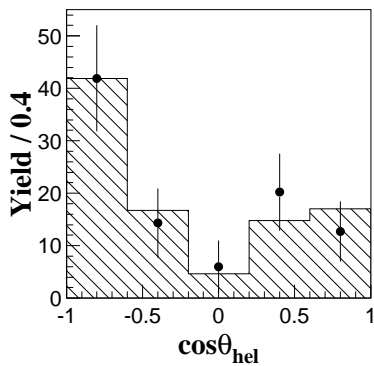}
\caption{Data points show the background-subtracted $\cos\theta_{\rm
    hel}$ distribution for the $\rho$; hatched histogram shows the
    distribution for signal MC. The asymmetry in these distributions is
    due to the $\pi^0$ momentum requirement.}
\label{datahel}
\end{center}
\end{figure}

To verify that the signal we observe is due to $B^+\to \rho^+\pi^0$
decay, we examine the helicity and $M(\pi^+\pi^0)$ distributions.
Figure~\ref{datahel} shows the helicity angle distribution for signal
yields determined from $M_{\rm bc}$-$\Delta E$ fits, which is
consistent with that for signal MC events. The distribution for
continuum is approximately flat. 

\begin{figure}[htbp]
\begin{center}
\includegraphics[width=5cm,clip]{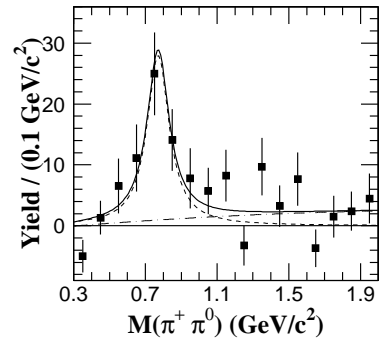}
\caption{Data points show the background-subtracted $M(\pi^+\pi^0)$
  distribution, the dashed (dot-dashed) line is for $\rho$ signal
  (non-resonant) component of the fit, the solid line is their sum.}
\label{datamrho}
\end{center}
\end{figure}

Figure~\ref{datamrho} shows the signal yields 
extracted from $M_{\rm bc}$-$\Delta E$ fits applied to individual
$M(\pi^+\pi^0)$ bins. A $\chi^2$ fit to the background subtracted
$M(\pi^+\pi^0)$ distribution is performed with a $\rho$ plus a
non-resonant $\pi\pi$ component included. 
The $\rho$ component is represented by a Breit-Wigner function with
mass and width fixed at their known values~\cite{PDG}.
The non-resonant $\pi\pi$ component is described by a second-order
polynomial with shape parameters determined from $B^+\to
\pi^+\pi^0\pi^0$ MC events, where the final state particles are
distributed uniformly over phase space.
The fit gives the fraction of non-resonant decays in the
$0.62{~\rm GeV}/c^2<M(\pi^+\pi^0)<0.92{~\rm GeV}/c^2$ $\rho$ signal
region as $(5.8\pm4.8)$\%. 
The non-resonant yield increased by $1\sigma$ is treated as a
systematic uncertainty.

We consider systematic errors in the branching fraction of the decay
$B^+\to \rho^+\pi^0$ that are caused by uncertainties in the
efficiencies of track finding, particle identification, $\pi^0$
reconstruction, continuum suppression, fitting and the possible
contribution from non-resonant decays.
We assign a 1.2\% error for the uncertainty in the tracking
efficiency. This uncertainty is obtained from a study of partially
reconstructed $D^*$ decays.
We also assign a $0.8$\% error for the particle identification
efficiency that is based on a study of kinematically selected $D^{*+}
\to D^0\pi^+$, $D^0\to K^-\pi^+$ decays;
an 8.0\% systematic error for the uncertainty in the two-$\pi^0$ detection
efficiency that is determined from data-MC comparisons of
$\eta\to\pi^0\pi^0\pi^0$ with $\eta \to\pi^+\pi^-\pi^0$ and $\eta \to
\gamma\gamma$;
a  5.1\% systematic error for continuum suppression that is estimated
from a study of $B^+ \to \overline{D}{}^0\pi^+$, and $\overline{D}{}^0\to
K^+\pi^-\pi^0$ decays;
a systematic error of $^{+4.3}_{-0.4}\%$ that is obtained from changes in
signal yields that occur when each parameter of the fitting functions
is varied by $\pm 1\sigma$;
a $^{+0}_{-10.5}\%$ systematic error to account for a possible contribution
from non-resonant decays.
Moreover, a $1\%$ error due to backgrounds from charmless $B$
decays other than $B\to \rho\rho$ and $B\to \pi^0\pi^0$ is
estimated by fitting the data with an additional component with the
yield fixed at the MC-expected value. The change in signal yield is
taken as a systematic error to account for this contamination.
We also include a 0.5\% error for the uncertainty in the number of
$B\bar{B}$ events in the data sample.
We obtain the branching fraction
$${\cal B}(B^+ \to \rho^+ \pi^0)=(13.2\pm 2.3({\rm
  stat.})^{+1.4}_{-1.9}({\rm sys.}))\times 10^{-6}.$$\\

Direct $CP$ violation would be indicated by an asymmetry in the partial
rates for $B^-\to \rho^-\pi^0$ and $B^+\to \rho^+\pi^0$:
$$
{\cal A}_{CP}
\equiv \frac{\Gamma{(B^- \to \rho^-\pi^0)}-\Gamma{(B^+\to
    \rho^+\pi^0)}}{\Gamma{(B^-\to \rho^-\pi^0)}+\Gamma{(B^+\to
    \rho^+\pi^0)}}.$$
The $B^\mp \to \rho^\mp\pi^0$ candidates are self-tagged, but this
tagging is provided by a single low-momentum charged pion that has
a relatively large combinatorial background.  This produces some
wrong-tagging, which results in a dilution of the true asymmetry:
${\cal A}_{\rm obs}=(1-2w){\cal A}_{CP}$, where ${\cal A}_{\rm obs}$
is the observed asymmetry and $w$ is the fraction of events that are
incorrectly tagged, determined  from  MC to be $w = 0.091\pm 0.015$.

We perform a simultaneous fit to extract the charge asymmetry by 
using the same components as shown in Fig.~\ref{fitdata_proj} and
introducing asymmetry parameters into the fit for the 
$B\to \rho\pi$ signal and also for the continuum and $b\to c$
backgrounds.
The  fit result is $A_{\rm obs}=0.05 \pm 0.16$, corresponding
to a $w$-corrected asymmetry ${\cal A}_{CP}=0.06\pm0.19$.

The charge symmetry of the detector performance and reconstruction
procedure is verified with a sample of $B^+\to \overline{D}{}^0 \pi^+$,
$\overline{D}{}^0\to K^+\pi^-\pi^0$ decays and their charge conjugates.
We apply the same procedure that is used for $B^+\to\rho^+\pi^0$ to
select $B^+\to\overline{D}{}^0\pi^+$ candidates and extract signal
yields by fitting the $\Delta E$ distribution.
For these events we find a direct-$CP$-violating asymmetry of
$-0.03\pm0.02$. We assign 0.03 as the systematic error
associated with detector and reconstruction effects. 
The systematic error associated with the fitting procedure is
determined to be 0.01 by shifting each fitting function parameter by
$\pm 1\sigma$ and taking the quadratic sum of the resulting changes in
${\cal A}_{CP}$. 
An error of $^{+0.00}_{-0.05}$ for non-resonant background is estimated
by subtracting the non-resonant component, obtained from the fit to
$M(\pi^-\pi^0)$ or $M(\pi^+\pi^0)$ and increased by $\pm 1\sigma$,
 from the $B^-$ or $B^+$ signal yields. The change in ${\cal
   A}_{CP}$ is taken as the systematic error. 
A 0.01 systematic error due to the background from other rare $B$
decays is assigned.
The quadratic sum of these errors is taken as the total systematic
error. We obtain a $CP$-violating charge asymmetry that is consistent
with zero:
$$
{\cal A}_{CP}(B^\mp\to \rho^\mp\pi^0) = 0.06\pm0.19({\rm
  stat.})^{+0.04}_{-0.06}({\rm sys.}).$$

In summary, we observe the decay $B^+ \to \rho^+ \pi^0$ with a
statistical significance of 8.1$\sigma$ and measure its branching
fraction.
The results are consistent with those from the BaBar
experiment~\cite{BaBar_rhoppi0}.
This measurement provides one of the essential quantities to constrain
$\phi_2$ from an isospin analysis of $B \to \rho \pi$ decays.
The measured charge asymmetry is consistent with zero.\\

\begin{acknowledgments}
We thank the KEKB group for the excellent operation of the
accelerator, the KEK Cryogenics group for the efficient operation of
the solenoid, and the KEK computer group and the NII for valuable
computing and Super-SINET network support.  We acknowledge support
from MEXT and JSPS (Japan); ARC and DEST (Australia); NSFC (contract
No.~10175071, China); DST (India); the BK21 program of MOEHRD and the
CHEP SRC program of KOSEF (Korea); KBN (contract No.~2P03B 01324,
Poland); MIST (Russia); MESS (Slovenia); NSC and MOE (Taiwan); and DOE
(USA).
\end{acknowledgments}


\begin{thebibliography}{99}
\bibitem{charge_conjugate} 
The inclusion of charge conjugate modes is implied unless stated otherwise.
\end{thebibliography}
\end{document}